\documentclass[aps,pra,reprint,amsmath,amssymb,superscriptaddress,longbibliography]{revtex4-2}

\usepackage{graphicx}

\usepackage{xcolor}
\usepackage{hyperref}
\hypersetup{
     colorlinks=true,
     linkcolor=blue,
     filecolor=blue,
     citecolor = black,      
     urlcolor=blue
     }
\usepackage{url}
\usepackage{enumitem} 

\def\TNS{Ta$_2$NiSe$_5$\,}

\def\TNSX{Ta$_2$Ni(Se$_{1-x}$S$_x$)$_5$\,}
\def\Tc{$T_c$\,}

\def\cm1{cm$^{-1}$\,}
\def\B2g{$B_{2g}$\,}

\usepackage{color,soul}

\begin{document}

	\title{
		Comment on "Direct observation of excitonic instability in \TNS" [arXiv:2007.08212 (2020)]
	}
	
	\author{G. Blumberg}
	\email{girsh@physics.rutgers.edu}
	\affiliation{Department of Physics and Astronomy, Rutgers University, Piscataway, NJ 08854, USA}
	
	\date{February 15, 2021}

\begin{abstract}
The origin of the second order phase transition at 328\,K in \TNS, a prominent candidate for direct gap excitonic insulator, remains under fervent debate. 
The driving force for the transition can be revealed by identification of the soft mode's origin that may be deducted from polarization resolved Raman scattering experiments. 
Such studies were recently reported in 
[\href{https://arxiv.org/pdf/2007.07344.pdf}{arXiv:2007.07344 (2020)}], 
[\href{https://arxiv.org/pdf/2102.07912.pdf}{arXiv:2102.07912 (2021)}], 
[\href{https://arxiv.org/pdf/2007.01723v3.pdf}{arXiv:2007.01723v3 (2021)}] and 
[\href{https://arxiv.org/pdf/2007.08212.pdf}{arXiv:2007.08212 (2020)}].  
In this Comment, it is shown that the parameters derived in a recent arXiv by Kwangrae Kim et. al. [\href{https://arxiv.org/pdf/2007.08212.pdf}{arXiv:2007.08212 (2020)}], including the Weiss temperature for excitonic transition, are based on inconsistent data.  
\end{abstract}
	
	\maketitle

\section{Introduction}	

The low-dimensional Ta$_2$NiSe$_5$ material is a prominent candidate for direct gap excitonic insulator (EI)~\cite{ARPES2009,ARPES2014,Transport2017}.
It is a layered crystal showing a semiconducting behavior below a second-order phase transition from high-temperature orthorhombic $Cmcm$ phase to a monoclinic $C2/c$ one at $T_c=328$\,K.   
The phase transition is associated with a small change in the angle between $a$ and $c$ crystallographic directions from 90$^{\circ}$ to 90.53$^{\circ}$ that breaks two of the mirror symmetries~\cite{Structure1986}. 
The transition is coupled to strain fields which results in the formation of 
domains below $T_c$ that were directly imaged by TEM~\cite{Structure1986,Mai2021}. 
Below $T_c$, a gap opens and an anomalous dispersion of the hole-like band detected by ARPES studies has been taken as an indication for the excitonic character of the transition~\cite{ARPES2009,Chen2020}. 

The group-subgroup relation between the $Cmcm$ and $C2/c$ space groups implies  freezing of a \B2g symmetry mode on cooling toward $T_c$. 
The origin of the soft mode determines the driving force of the transition, i.e., 
for a true EI the condensation of \B2g symmetry excitons is expected below $T_c$~\cite{Pavel2020,Kim2020,KaiserErr}. 
Nevertheless, the changes in spectral and transport properties below $T_c$ could also originate from modifications of the lattice structure at $T_c$. 
While Ta$_2$NiSe$_5$ has been actively investigated~\cite{ARPES2014,Transport2017,IR2017,IR2018,IR2018a,millis2019}, a purely structural origin of the transition has not been ruled out~\cite{ARPES2020,subedi2020,baldini2020}. 
Indeed, on one hand, softening of the \B2g symmetry acoustic mode has been directly observed by an inelastic x-ray scattering experiment~\cite{XRay2018}, 
and softening of the \B2g symmetry optical phonon mode has been predicted by DFT calculation~\cite{subedi2020}. 
On the other hand, no direct observation of the behavior of all the relevant modes have been reported. 
All excitations of the \B2g symmetry: excitons, phonons and strain fields, - could contribute to the transition, further complicated by the presence of electronic-lattice interaction. 
Thus, the identification of the soft mode, the main driver of the transition, is of fundamental importance.

The questions above can be addressed directly by low-frequency polarization resolved Raman spectroscopy that probes the excitations of the system by inelastic two-photon process~\cite{Pavel2020}. 
Applied to Ta$_2$NiSe$_5$ above $T_c$, the Raman response in $ac$ polarization geometry probes excitations with the symmetry of $ac$-type quadrupole, the same as that of the \B2g symmetry order parameter (OP), allowing direct observation of the soft mode.  
The character of the soft mode reveals the origin of the transition. 
If the transition is structural only, an optical phonon, represented by a sharp spectral peak, softens to zero energy at $T_c$. 
On the contrary, for an excitonic transition in a semimetal, critical fluctuations have a broad relaxational lineshape due to the Landau damping and are enhanced at low frequencies close to $T_c$. 
Finally, a purely strain-driven (ferroelastic transition) is not expected to lead to an observable soft mode in a Raman experiment \cite{gallais2016}.

Recently, polarization-resolved Raman spectroscopy was employed to study the origin of the transition in Ta$_2$NiSe$_5$~\cite{Pavel2020,Kim2020,KaiserErr} and in \TNSX family of EI candidate materials~\cite{Mai2021}. In Refs.~\cite{Pavel2020,Mai2021} the authors demonstrate both, 
(i) the critical nature of excitonic fluctuations and 
(ii) the hybridization origin of the band gap that develops below $T_c$;  
thus, prove that the condensation of excitons is the chief reason for the transition. 
Furthermore, strong coupling of the excitons to the strain fields and to two low energy \B2g symmetry optical phonons was confirmed in these studies. 

The critical nature of excitonic fluctuations has also been claimed by recent studies of K.\,Kim et. al.~\cite{Kim2020}, however, the presented data and the details of analyses are distinct from the studies~\cite{Pavel2020,Mai2021}.
In this Comment, it is shown that the discrepancies stem from artifacts of measurements and erroneous analyses applied in~\cite{Kim2020}.

\section{Experimental}	

For the low-frequency Raman studies~\cite{Pavel2020,Mai2021} 
authors have employed a custom fast $f$/4 high resolution 500/500/660\,mm focal lengths triple-grating spectrometer with 1800\,mm$^{-1}$ master holographic gratings comprised of 
(i) aberration corrected subtractive stage providing a reliable about 11 orders-of-magnitude stray light rejection at as low as about 4\,\cm1 from the elastic line, 
(ii) a third stage monochromator, and (iii) a liquid-nitrogen-cooled charge-coupled device (CCD) detector (Princeton Instruments) for the scattered light acquisition. 
For polarization optics, we used a Glan-Taylor polarizing prism (Melles Griot) with a better than 10$^{-5}$ extinction ratio to clean the laser excitation beam and a  broad-band 50\,mm polarizing cube (Karl Lambrecht Corporation) with an extinction ratio better than 1:500 for the analyzer. 
All the acquired data were corrected for the spectral response of the spectrometer~\cite{Mai2021}. 

\begin{figure}[b]
\includegraphics[width=0.8\linewidth]{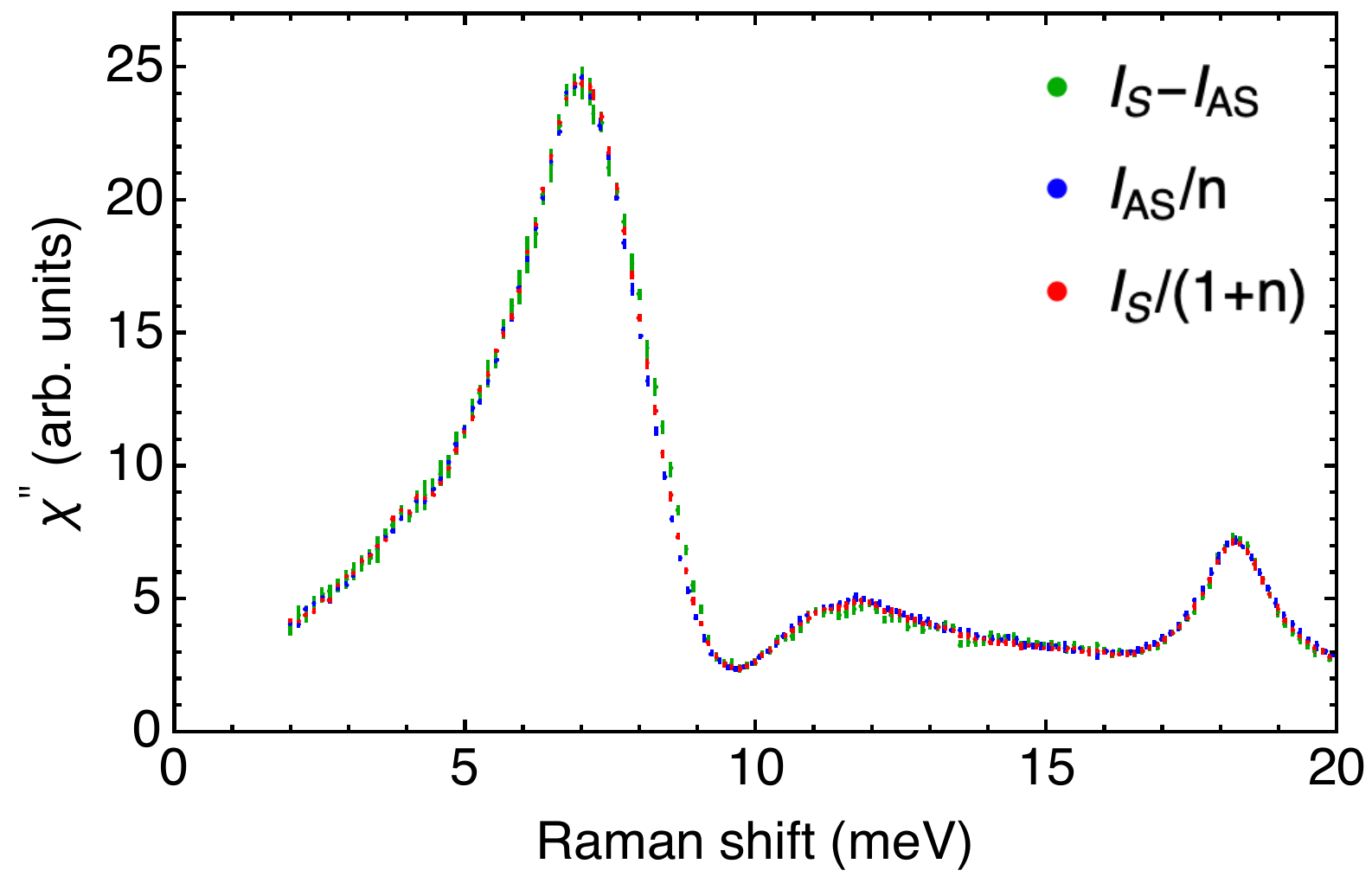}
\caption{ 
The comparison of Raman response function $\chi^{\prime\prime}(\omega,T)$ calculated from the Stokes intensity $I_S/(1+n(\omega,T))$ (red), the anti-Stokes intensity $I_{AS}/n(\omega,T)$ (blue), and by subtraction $I_{AS} - I_S$ (green) measured with 56\,mW laser power focused into 50$\times$100\,$\mu$m$^{2}$ excitation spot (power density 1.4\,kW/cm$^2$) at 295\,K environmental temperature. 
The Bose factor is denoted by $n(\omega,T)$.
The detailed balance analyses derives same 371$\pm$7\,K equilibrium temperature for both, the electronic continuum and the lattice optical phonon modes~\cite{Mai2021}. 
The error bars shown are one standard deviation. 
}
\label{fig:StAst}
\end{figure}

The freshly cleaved samples were examined for strain-free area under Nomarski microscope before loading in a custom continuous helium-gas-flow cryostat. 
The Raman-scattering measurements were performed in a quasi-back-scattering geometry. 
The 647.1\,nm line from Kr$^+$ ion laser was used for excitation. 
The incident light was focused to an elongated 50$\times$100\,$\mu$m$^{2}$ spot. 
For all data taken below 310\,K, laser power of 8\,mW was used (power density 200\,W/cm$^2$); 
and up to seven times higher laser power was used for measurements above 310\,K. 
All reported data were corrected for laser heating in the measurement spot that has been determined by two  mutually consistent ways: 
(i) by Stokes/anti-Stokes intensity ratio analysis, based on the principle of detailed balance, Fig.\,\ref{fig:StAst}; and 
(ii) by checking laser power that is inducing the phase transition~\cite{Mai2021}.  

\begin{figure}[b]
\includegraphics[width=0.8\linewidth]{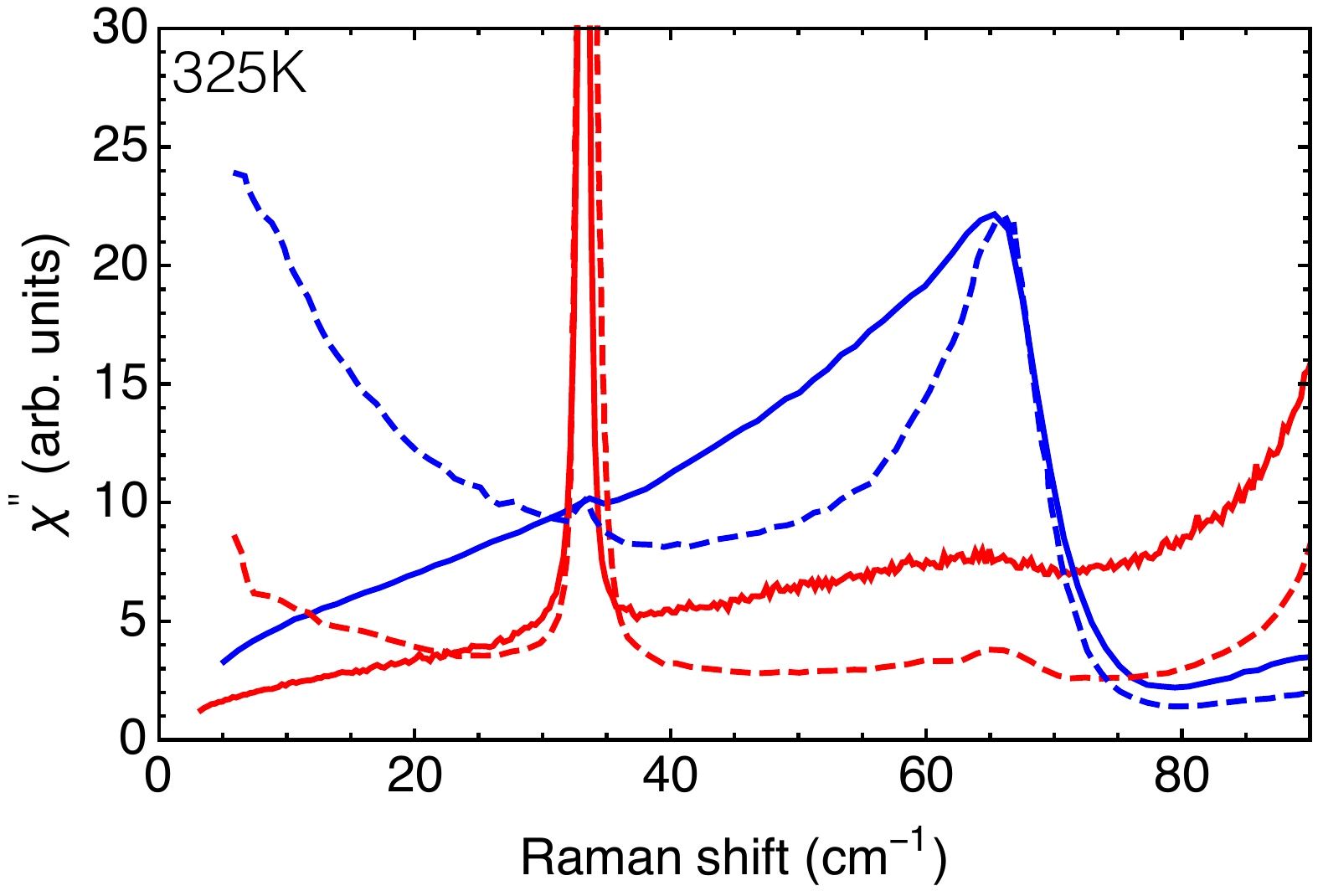}
\caption{The comparison of Raman response data $\chi^{\prime\prime}(\omega)$ at 325\.K, 
measured with aberration corrected custom triple grating spectrometer (solid lines)~\cite{Pavel2020,Mai2021} to the data adapted from Fig.\,3(a) in Ref.\,\cite{Kim2020,Kim2021PC} (dashed lines) for $ac$ polarization (blue) and $aa$ polarization (red). 
}
\label{fig:Kim2}
\end{figure}

In the study by Kwangrae Kim et. al.~\cite{Kim2020} the authors used a home-built setup with insufficient information disclosed. 
Based on the comparison data, see Fig.\,\ref{fig:Kim2}, we challenge the statement that the setup allows investigation of low-energy signals without a contamination. 
The authors do not disclose if the data has been corrected to the spectral response of spectrometer. 
The red HeNe laser line with 1.85\,mW power focused to 2\,$\mu$m spot is claimed to be used for excitation (power density 60\,kW/cm$^2$ -- more than two orders of magnitude larger than in Refs.~\cite{Pavel2020,Mai2021}). 
The authors do not disclose how crystal cooling and heating was achieved, other than the samples were mounted in an open-cycle cryostat (Oxford Instruments), 
nor do they explain how was the temperature in the laser spot determined~\footnote{We estimate, based on thermo-conductivity model calculation, that for 1.85\,mW laser power focused into a 2\,$\mu$m diameter spot the laser heating should be around 35 times larger than in \cite{Pavel2020,Mai2021}, which may result in a distortion of the data due to, e.g. inhomogeneous heating or sample damage; however, no proper analysis of these effects is reported in~\cite{Kim2020}.
}.

\section{Discussion of the data and the analyses in Kwangrae Kim et. al.} 

\begin{figure}[t]
\includegraphics[width=0.8\linewidth]{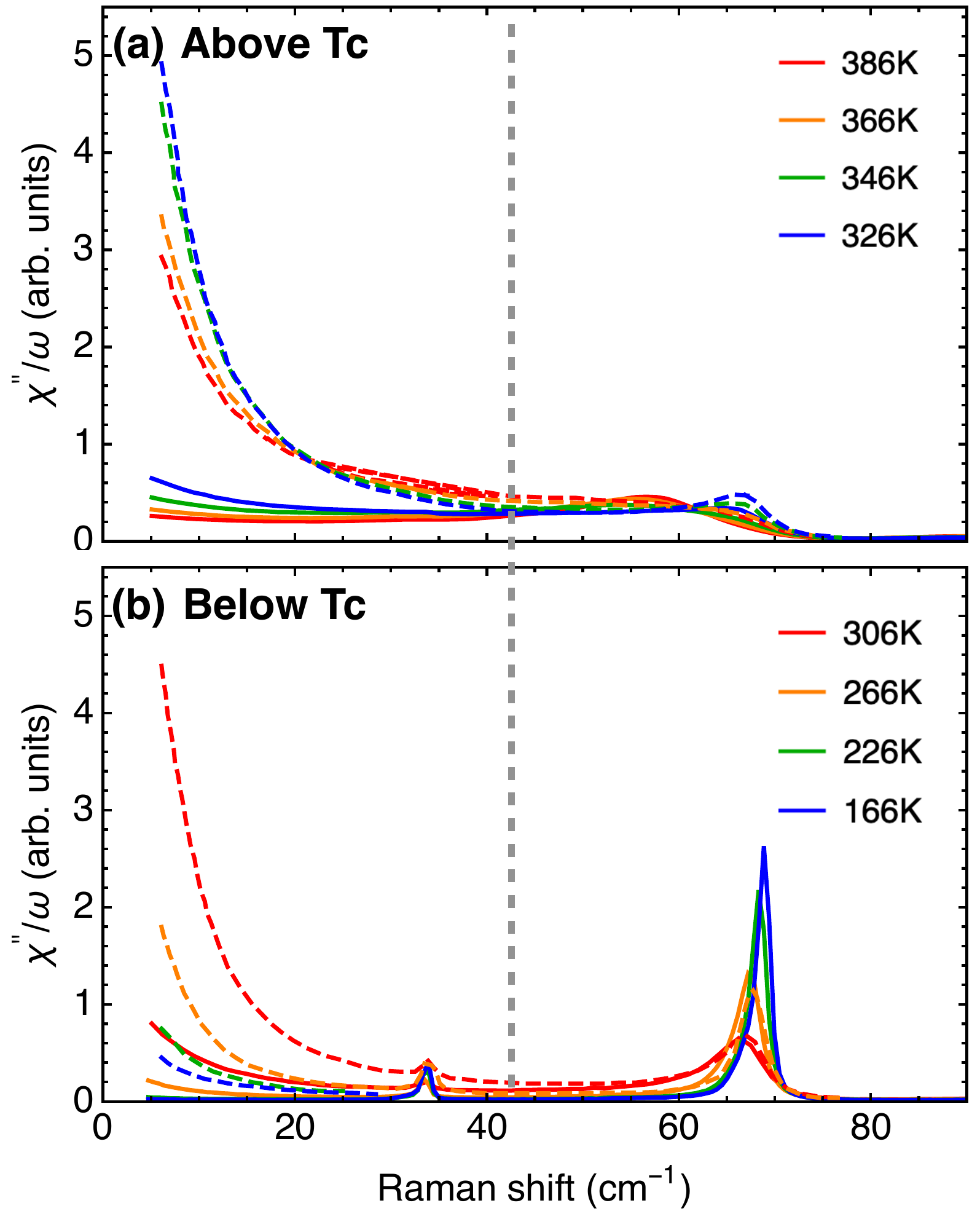}
\caption{The comparison of Raman response data divided by frequency, 
$\chi^{\prime\prime}(\omega,T)/\omega$, 
measured with aberration corrected custom triple grating spectrometer (solid lines)~\cite{Pavel2020,Mai2021} to the data adapted from Fig.\,2(a) and (b) in Ref.\,\cite{Kim2020}. 
(a) Data above \Tc, and (b) data below \Tc. 
The vertical gray dashed line indicates Raman shift below which the Ref.\,\cite{Kim2020} data is likely influenced by stray light or other contamination signals. 
}
\label{fig:Kim}
\end{figure}
While the general philosophy of the analyses in Ref.\,\cite{Kim2020} is somewhat similar to Ref.~\cite{Pavel2020}, the data and deduced parameters are sharply distinct. 
The difference is obvious from direct comparison of the Raman response functions, Fig.\,\ref{fig:Kim2}-\ref{fig:Kim}. 
Therefore, the following comments are in order: 
\begin{enumerate}[label = (\roman*), left=-0.25\parindent, itemsep=0pt, parsep=0pt, topsep=4pt]
\item 
Direct comparison of the $\chi^{\prime\prime}(\omega,T)$ response and $\chi^{\prime\prime}(\omega,T)/\omega$ function derived from the Raman data measured with aberration corrected custom triple grating spectrometer~\cite{Pavel2020,Mai2021} to the data from Fig.\,2a-b and Fig.\,3a in Ref.\,\cite{Kim2020} (dashed lines) indicate a significant discrepancy below 60\,\cm1. 
Furthermore, the upturn in Raman response function $\chi^{\prime\prime}(\omega,T)$, which by definition is an odd function of the Raman shift $\omega$, down to the lowest measured 5\,\cm1 is unphysical. 
The reason for such unphysical signal upturn is difficult to address here as the experimental description is insufficiently presented in~\cite{Kim2020}. 
Thus, the derived static Raman susceptibility $\chi(\omega=0,T)$ shown in Fig.\,2 of Ref.\,\cite{Kim2020} cannot be reliable. 
\item
Furthermore, before the Kramers-Kronig transformation, the $\chi^{\prime\prime}(\omega,T)/\omega$ must be continued to zero frequency, which has not been done in~\cite{Kim2020}, introducing further errors into analyses of $\chi(\omega=0,T)$ and hence erroneous determination of the Weiss temperature. 
In particular, because $\chi^{\prime\prime}(\omega,T)$ erroneously does not approach zero in the zero-frequency limit (see Fig.\,\ref{fig:Kim2}, dashed lines), the integral in the Kramers-Kronig transformation is divergent at low frequencies, which makes it dependent on the lower cutoff and thus unreliable.
\item
The use of non-interacting model in the treatment of the data in Fig.\,3 of Ref.\,\cite{Kim2020}, where data is decomposed into a simple sum of 'the mode 2' (blue) and 'a continuum' (red) is improper as such treatment neglects (a) the interference effects and (b) modes' renormalization due to the interaction between them. 
Instead, the proper complete Fano analyses (not just na\"ive Fano shaped 'mode 2') must be applied~\cite{Mai2021}. 
\item
The description of the model used for interpretation is insufficient, and no physical justification is offered for the temperature dependencies of the deduced parameters displayed in Figs.\,2c-e and 3b. 
Why would 
(a) the amplitude of both phononic and electronic responses, and  
(b) electron-phonon coupling constant be strongly temperature dependent? 
(c) How can phononic relaxation rate rapidly reduce upon heating above 400K? 
(d) How can electronic relaxation rate be much smaller than temperature within a Drude model? 
\end{enumerate}
\medskip

\section{Conclusions}

We have shown in this Comment that while the philosophy of the data analyses presented by Kwangrae Kim et. al.~\cite{Kim2020} is similar to the one reported earlier~\cite{Pavel2020}, inconsistencies of the Raman data, the failure to continue Raman response to zero frequency limit, erroneous data decomposition that neglects the renormalization due to exciton-phonon interaction and the interference terms in the spectra result in unphysical temperature dependencies of deduced parameters. 

\begin{acknowledgments}
I acknowledge discussion with Mai Ye and Pavel Volkov, and the collaboration with the Technion group: Himanshu Lohani, Irena Feldman and Amit Kanigel~\cite{Pavel2020,Mai2021}.  
The work was supported by NSF Grant No. DMR-1709161. 

\end{acknowledgments}

\end{document}